\documentclass[aps,prc,twocolumn,groupedaddress]{revtex4-1}
\usepackage{graphicx}% Include figure files
\usepackage{amssymb,amsmath}
\usepackage{multirow}
\usepackage{xcolor}
\usepackage{caption}
\usepackage{graphicx,subcaption}
\usepackage{floatrow}
\usepackage{makecell}
\begin{document}

%\preprint{}

\title{Uncovering the nature of low-lying dipole states with QRPA calculations: is Z=42 the answer?}% Force line breaks with \\


\author{E. J. In$^{1}$}
\email{in2@llnl.gov}
\author{E. V. Chimanski$^{2}$}
\author{J. E. Escher $^{1}$}
\author{S. P{\'e}ru$^{3,4}$}
\author{A. Thapa$^{1}$}
\author{W. Younes$^{1}$}
%\thanks{}

\affiliation{%
 Lawrence Livermore National Laboratory, Livermore, 94550, CA, USA $^{1}$ \\
 National Nuclear Data Center, Brookhaven National Laboratory, Upton, 11973, NY, USA $^{2}$ \\ 
 CEA, DAM, DIF, Arpajon, France $^{3}$ \\ 
 Universit\'e Paris-Saclay, CEA, LMCE, 91680 Bruy\`eres-le-Ch\^atel, France $^{4}$ \\
% This line break forced with \textbackslash\textbackslashhttps://www.overleaf.com/project/620a9135037a08431f0cd822
}%

\date{\today}

\begin{abstract}

The pygmy dipole resonance (PDR), marked by enhanced electric dipole strength near particle emission energies, offers a unique perspective on the collective dynamics of nuclear structure. Its precise nature, particularly its degree of collectivity, remains a topic of debate.
In this study, we investigate low-energy dipole excitations in spherical Mo isotopes ($^{82}$Mo to $^{98}$Mo) using a fully consistent Hartree-Fock-Bogoliubov (HFB) and quasiparticle random phase approximation (QRPA) framework. 
We observe that an enhancement in dipole strength near particle emission energies is closely correlated with the development of either neutron or proton skins. 
To further understand the nature of this enhancement, we examine the behavior of proton and neutron transition densities. 
Our analysis shows that these (low-lying dipole) states exhibit distinct characteristics involving in-phase oscillations within the nucleus and neutron- or proton-dominated oscillations at the surface, while the primary contributor to this enhancement displays an intricate underlying structure.
%Transition density analysis indicates that $E1$ excitations near the particle emission energies exhibit characteristics distinct from the isovector giant dipole resonance (IVGDR), involving in-phase oscillations within the nucleus and neutron- or proton-dominated oscillations beyond the surface. On the other hand, the primary contributor to this enhancement displays a more intricate underlying structure. {\color{blue} (meaning unclear) We assess contributions of the state with different characteristics to the overall magnitude of the enhancement.}
We also investigate the collectivity of these excitations by analyzing two-quasiparticle fragmentations and relative energy shifts. Our findings reveal that skin oscillation states exhibit moderate collectivity, as indicated by substantial configuration mixing, but limited coherence, whereas the GDR states exhibit strong coherence and large energy shifts characteristic of fully developed collective motion.
This study paves the way for future investigations into the collective nature of low-energy dipole states in the enhancement region, particularly in deformed nuclei, where nuclear shape effects may play a crucial role in their excitation dynamics.
%This study establishes a foundation for further investigations into the collective nature of low-energy dipole states in the enhancement, particularly in deformed nuclei, where nuclear shape effects may play a crucial role in their excitation dynamics.
%These findings pave the way for future studies into the collective behavior of PDRs in deformed nuclei.

\end{abstract}


\maketitle


\clearpage
\section{Introduction}\label{intro}

The study of nuclei with neutron excess has drawn considerable attention in nuclear structure physics, particularly due to the development of neutron skins \cite{Paar2007, Savran2013, Bracco2019, Lanza2023}. Neutron skins arise from the spatial decoupling of an extended neutron distribution from the core nucleons and can give rise to novel excitation modes.
One such excitation is the so-called pygmy dipole resonance (PDR), characterized by an enhancement of electric dipole strength near the neutron separation threshold ($S_n$). The PDR is often interpreted as a signature of neutron skin oscillation and has been typically observed in nuclei with a neutron excess.
While the PDR accounts for only a small fraction of the total energy-weighted sum rule (EWSR), it is of particular interest because it is expected to reflect the motion of the neutron skin against the core, which is closely linked to the isospin dynamics and the symmetry energy of nuclear matter~\cite{Piekarewicz2006, Andrea2010}.
In order to obtain more evidence for such a peculiar mode, studies of low-energy dipole excitations over sufficiently long isotopic chains are essential.
In addition to its structural significance, the PDR plays a crucial role in nuclear reactions, particularly in astrophysical nucleosynthesis processes \cite{Goriely1998, Goriely2004}. The presence of enhanced low-energy $E1$ strength impacts neutron capture rates in the $s$- and $r$-processes, influencing the synthesis of heavy elements.  
For example, PDR contributions in $^{206}$Pb affect the $^{205}$Pb(n,$\gamma$)$^{206}$Pb reaction cross section \cite{Tonchev2017}, a key contributor to the astrophysical $s$-process where $^{205}$Pb acts as a branching point nucleus.

Recent advances in radioactive beam techniques have provided compelling evidence for these exotic excitation modes in neutron-rich nuclei \cite{Adrich2005, Savran2013, Bracco2019, Brown2025}. Theoretically, various approaches including the Hartree-Fock (HF) plus random phase approximation (RPA) \cite{Adams1996, Catara1997, Lanza2009, Lanza2011}, Hartree-Fock-Bogoliubov (HFB) plus the quasiparticle RPA (QRPA) \cite{Martini2011}, the quasiparticle-phonon model (QPM) \cite{Tsoneva2008}, and relativistic approaches \cite{Vretenar2001, Paar2005PLB-energy, Piekarewicz2006, Argeaga2009} have revealed strong correlations between PDR properties and neutron skin thickness.
Despite extensive theoretical and experimental investigations, the collective nature of the PDR remains a subject of debate.
%{\color{blue} (Sophie: Should we give a number divided by A in order to fairly compare the collectivity of the heavier nuclei with the lighter ones?)}
In light nuclei, low-lying dipole excitations are typically non-collective, driven by loosely bound single-neutron transitions. In contrast, in medium-mass and heavy neutron-rich nuclei, some low-lying dipole states exhibit a more distributed particle-hole structure, suggesting a certain degree of collectivity \cite{Vretenar2001NPA, Paar2007, Paar2009}.
While relativistic RPA and QRPA calculations predict collective characteristics in certain PDR states, alternative models such as QRPA with phonon coupling indicate that PDR states consist of only a few neutron particle-hole configurations, challenging their interpretation as a coherent collective mode \cite{Colo2001,Tsoneva2008}. Other studies such as Ref. \cite{Lanza2009} have shown that while the PDR involves multiple configurations, destructive interference among them can prevent it from achieving transition strengths comparable to a strongly collective mode such as the GDR.
While most investigations have focused on neutron-rich nuclei, a few theoretical predictions indicate that proton-rich nuclei may also exhibit proton PDRs \cite{Martini2011, Paar2005, Paar2005PLB}. However, due to the Coulomb barrier, significant proton skins are less likely to form in medium-to-heavy nuclei, making proton PDRs challenging to observe. As a result, their structural characteristics and potential astrophysical relevance remain largely unexplored.

%Savran2013: In many models, the E1 strength located at low excitation energies shows the signature of a neutron-skin oscillation. In neutron deificient nuclei a PDR correlated to a proton skin has been predicted. While the low-lying component of the E1 strength is observed in almost all calculations, the degree of collectivity is under debate.
%Lanza2009: Instead, destructive interference among these configurations results in transition strengths comparable to those of a single particle-hole excitation, indicating a limited degree of collectivity.
%2009PRC_Lanza: For very light nuclei (halo nuclei), the soft modes are not collective states, but rather they are generated by the radial extension of the very weakly bound single-particle neutron states.
%their contributions partially cancel out, resulting in a reduced transition probability comparable to that of a single particle-hole excitation rather than a strongly collective mode like GDR
%while PDR states involve multiple configurations, their contributions do not add up coherently as in strongly collective excitations like the giant dipole resonance (GDR). This reinforces the notion that the PDR exhibits a more complex nature, with its degree of collectivity remaining an open question.

Molybdenum (Mo) isotopes serve as an important laboratory for PDR studies due to their involvement in multiple nucleosynthesis processes ($r$-, $s$-, and $p$-processes) \cite{Wieser2007,Stephan2019}. Experimentally, enhanced dipole strength near the $S_n$ threshold has been observed in Mo isotopes, indicating the presence of PDRs. 
%Photon scattering experiments on $^{92}$Mo have identified possible pygmy resonances below the $S_n$ threshold \cite{Rusev2006}. 
Studies of stable even-even isotopes from $^{92}$Mo to $^{100}$Mo with ($\gamma, xn$) studies report enhanced electric dipole strength below and above the $S_n$ threshold, attributed to deformation effects and supported by microscopic QRPA calculations~\cite{Rusev2006,Rusev2009,Erhard2010}. Additional evidence from $(\alpha,\alpha'\gamma)$ scattering on $^{94}$Mo confirms the PDR in the 6-10 MeV range \cite{Derya2013}, while nuclear resonance fluorescence experiments observe enhanced dipole structures between 5.5 and 7.5 MeV \cite{Roming2013}. 
Despite these experimental efforts, theoretical studies of the PDRs in Mo isotopes remain limited \cite{Heim2021, Pascu2012, In2025}.
While previous studies examining the $\gamma$-ray strength function ($\gamma$-SF) in $^{94}$Mo via the $^{93}$Nb($p$,$\gamma$)$^{94}$Mo reaction using the Gogny D1M+QRPA $\gamma$-SF model \cite{Heim2021}, and interacting boson approximation (IBA) calculations have predicted individual $E1$ transitions in the 5–7 MeV range \cite{Pascu2012},
a comprehensive understanding of the PDR in Mo isotopes-including its location, collectivity, and dependence on neutron and proton skins-remains an open question.

To gain deeper insight into the nature of low-energy dipole excitations, detailed studies across a reasonably extensive isotopic chains are necessary. In this work, we present a comprehensive analysis of dipole strength enhancements in spherical even-even Mo isotopes ($^{82-98}$Mo).
Our study employs the HFB+QRPA framework with the Gogny D1M parametrization \cite{Gogny1980, Goriely2009}, chosen for its improved accuracy in nuclear mass predictions and its superior performance in describing nuclei with neutron excess, making it well-suited for this investigation.
The primary focus of this work is to disentangle the microscopic structure underlying the low-energy dipole enhancement, by identifying contributions from skin oscillation states and major low-energy peaks, and investigate the evolution of these excitations in relation to skin development.

\section{Methods}

Low-lying and giant resonance collective vibrational states in nuclei, both near and far from the stability line, have been successfully described using mean-field-based approaches \cite{Paar2007, Sophie2014, Lanza2023}. This section outlines our approach, which employs the QRPA built on top of HFB calculations. For a detailed explanation of these approaches, readers are referred to the literature \cite{Ring1980, Walid2009, Sophie2014, Walid2019, Emanuel2024}.

\subsection{HFB+QRPA}\label{HFBQRPA}
 
The HFB equations are solved in the harmonic-oscillator (HO) basis in cylindrical coordinates. The calculations for even-even nuclei with particle number constraints assume axial symmetry of the nucleus and impose time-reversal invariance. Under these assumptions, the projection $K$ of the angular momentum $J$ on the symmetry axis is a  good quantum number. In addition, we impose the parity symmetry $\pi$.
 The quasiparticle operators $\eta^{\dagger}$ and the HO particle operators $c^{\dagger}$ and $c$ are related through:
\begin{equation} \label{bogoliubov}
\eta_i^{\dagger} = \sum_{\alpha} \left[ u_{i\alpha} c_\alpha^{\dagger} - v_{i\alpha} c_\alpha \right], 
\end{equation}
where the $u$ and $v$ are Bogoliubov transformation matrices. 
The QRPA matrix is expressed in terms of two-quasiparticle (2qp) states obtained from the HFB solutions at the minimum of the axially-symmetric deformed potential energy curves. To ensure consistency, the same Gogny D1M parametrization is employed in both the HFB and the QRPA calculations. All the 2qp configurations are included without imposing an excitation energy cutoff. Excitations beyond 2qp configurations are not considered. The Coulomb terms are treated exactly in both HFB and QRPA calculations. In a fully consistent calculation, spurious center-of-mass states are properly separated from the physical spectrum, appearing near zero energy in the calculations.

QRPA calculations are performed separately for each $K^{\pi}$. The QRPA eigenstates, $|w\rangle$, corresponding to the $w$-th eigenstate with a given $K^{\pi}$, are expressed as:
%The 2qp QRPA eigenstates, $|w\rangle \equiv |n,K^{\pi} \rangle$, corresponding to the n-th eigenstate with a given $K^{\pi}$, are expressed as:
\begin{equation}
|w\rangle = \theta_w^{\dagger} |0_{\mathrm{def}}\rangle,
\end{equation}
where $|0_{\mathrm{def}} \rangle$ is the correlated ground state and $\theta_w^{\dagger}$ is the QRPA excitation operator %with energies $\epsilon_w$,
\begin{equation} \label{QRPAope}
\theta_{w}^{\dagger} = \sum_{i<j} \left( X^w_{ij} \eta_i^{\dagger}\eta_j^{\dagger} - Y^w_{ij} \eta_j\eta_i \right).
\end{equation}
Here, the amplitudes $X$ and $Y$ are solutions of the QRPA matrix equations. The qp indices $i$ and $j$ are paired such that $K=k_i+k_j$ and $\pi=\pi_i\pi_j$. The resulting QRPA solutions represent superpositions of 2qp configurations with the same $K^{\pi}$. When these configurations sum up coherently, they correspond to collective states. 
%{\color{blue} (remove) For axially-symmetric deformed systems, calculations are performed in the intrinsic frame. Transformation to the laboratory frame requires the Wigner rotation matrices \cite{Edmonds1996}.} 
In spherical nuclei, the energy degeneracy of all $K$ components allows response functions for any multipolarity to be obtained from $K^{\pi} = 0^{\pm}$ results alone.

\subsection{Dipole excitations and transition densities}

%{\color{blue} Emanuel:  Maybe if don’t use these operates  (Eq. 4 and Eq. 5) explicitly I would go with Eq 6 and 7 only. I sometimes get confused with IV. Alternatively you can say before Eq 6, that using operators 4 and 4 we get Eq 6 and 7 ( also thinking out loud maybe) }

The electric dipole response is computed for both the charge isovector (IV) and mass isoscalar (IS) dipole operators, defined as:
%{\color{blue} need to verify sign}
\begin{align}
\hat{D}_{IV} &= \frac{eZ}{A} \sum_{n=1}^{N} r_n Y_{1m}(\hat{r}_n) - \frac{eN}{A} \sum_{p=1}^{Z} r_p Y_{1m}(\hat{r}_p), \label{IVoperator} \\
\hat{D}_{IS} &= \sum_{i=1}^{A} \left(r_i^3 - \frac{5}{3} \left<r^2\right> r_i \right) Y_{1m}(\hat{r}_i), \label{ISoperator}
\end{align}
where the $p$ and $n$ subscripts denote protons and neutrons, respectively. These operators include corrections to restore translational invariance.
The $IV$ operator probes out of phase motion between protons and neutrons, while the $IS$ operator is sensitive to in phase collective motion of nucleons.
%where $\left<r^2\right>$ denotes the ground state expectation value, and the inclusion of the second term in the operator ensurs that the strength distribution does not contain spurious components htat correspond to the centre-of-mass motion.

The response to the dipole excitation operator is expressed in terms of the reduced transition probability from the ground state to an excited state $|w\rangle$: 
\begin{align}
B(E1) &= \left| \int r^3 dr \left[ \frac{Z}{A} \delta \rho_{n}^{w}(r) - \frac{N}{A} \delta \rho_{p}^{w}(r) \right] \right|^2, \label{BIV} \\
B(IS) &= \left| \int r^3 \left(r^2-\frac{5}{3} \left<r^2\right> \right) dr \left[ \delta \rho_{n}^{w}(r) + \delta \rho_{p}^{w}(r) \right] \right|^2, \label{BIS}
\end{align}
where $\delta \rho_{n}^{w}(r)$ and $\delta \rho_{p}^{w}(r)$ are the neutron and proton dipole radial transition densities for the excited state $|w\rangle$, respectively. 
Equations~\eqref{BIV} and~\eqref{BIS} are obtained from the operators in Eqs.~\eqref{IVoperator} and~\eqref{ISoperator}, respectively.

Using the orthogonal set of spatial single particle wave functions $\phi_\beta({\bf{r}})$,
the intrinsic transition density $\rho_{\mathrm{tr}}^{w}(\bf{r})$ is expressed in second quantization as
\begin{equation}
\rho_{\mathrm{tr}}^{w}({\bf{r}}) = \sum_{\alpha\beta} \phi_\alpha^*({\bf{r}}) \phi_\beta({\bf{r}}) \langle w|c_\alpha^\dag  c_\beta|0_{\mathrm{def}} \rangle
\end{equation}
where the $\phi_\beta({\bf{r}})$ functions are defined in the cylindrical HO basis.
These transition densities are evaluated in the QRPA using the Bogolyubov transformation in Eq.~\eqref{bogoliubov}, the QRPA excitation operator in Eq.~\eqref{QRPAope} and the quasi-Boson approximation (QBA): 
%{\color{blue} I need to check this equation with Eq. (\ref{bogoliubov})}
\begin{align}
\rho_{\mathrm{tr}}^{w}({\bf{r}}) = \sum_{\alpha\beta} \phi_\alpha^*({\bf{r}}) \phi_\beta({\bf{r}}) 
& \sum_{ij} \left[ X^w_{ij}(u_{\alpha i}v_{\beta j} -u_{\alpha j} v_{\beta j}) \right. \nonumber \\
&\left. - Y^w_{ij}(v_{\alpha j} u_{\beta i} - v_{\alpha i} u_{\beta j}) \right].
\end{align}
The radial transition densities are expanded in terms of spherical harmonics $Y_{JK}(\Omega)$:
\begin{equation}
\delta \rho_{q}^{w}(r) = \int d \Omega \, Y_{JK}(\Omega) \rho_{\mathrm{tr},q}^{w}({\bf{r}})
\end{equation}
where the subscript $q$ refers to neutron ($q=n$) or proton ($q=p$). 
Radial transition densities provide critical insight into the nature of nuclear excited states by revealing how neutron and proton densities oscillate during excitation-either in-phase (IS), out-of-phase (IV), or exhibiting more complex behavior. Moreover, they are linked to radial form factors, which, in turn, affect reaction cross sections. Microscopic approaches for describing nuclear scattering processes based on nuclear structure calculations have been developed without relying on experimental input. Among these the Jeukenne-Lejeune-Mahaux (JLM) model has been widely used \cite{Bauge1998, Bauge2001, Dupuis2019, Aaina2024}. Transition densities serve as key ingredients for constructing transition potentials used in Distorted Wave Born Approximation (DWBA) or coupled-channels calculations, making their precise understanding essential for interpreting experimental data.

%The neutron and proton transition densities for each excited state are obtained by summing over the 2qp configurations for each nucleon type. From these, the IS and IV transition densities are constructed as: %{\color{blue} verify equations}
%\begin{eqnarray}
%\delta \rho_{E1}^{w}(r) = \delta \rho_{n}^{w} (r) - \delta\rho_p^{w} (r), \\
%\delta \rho_{IS}^{w}(r) = \delta \rho_{n}^{w} (r) + \delta\rho_p^{w} (r),
%\end{eqnarray}
%where $\delta \rho_{E1}^{w}(r)$ represents the IV transition density, with neutrons and protons oscillating out-of-phase, and $\delta \rho_{IS}^{w}(r)$ corresponds to the isoscalar (IS) transition density, where neutrons and protons oscillate in-phase. 

%The dipole photo absorption cross section of an excited nucleus is expressed as follows
%\begin{equation}
%\sigma_{E1}(E) = \frac{4\pi^2e^2}{\hbar c} (w_i-w_0) \sum |\left\langle \Psi_i |M| \Psi_0 \right\rangle|^2 \delta(w_i-w_0)
%\end{equation}
%where $M$ is the reduced transition operator and $\delta(w_i-w_0)$ is Lorentz function with the width $\Gamma$ (parameter) given as
%\begin{equation}
%\delta(w_i-w_0) = \frac{1}{2\pi} \, \frac{\Gamma}{(w_i-w_0)^2+(\Gamma/2)^2}.
%\end{equation}

\subsection{Quantification of collectivity} \label{collectivity}

Within the HFB+QRPA framework, we quantify the collectivity of dipole excitations by examining two key characteristics of the wave function: the spread over 2qp configurations (fragmentation number) and coherence among them.
An ideal collective state corresponds to the coherent superposition of many 2qp configurations, each contributing constructively with similar weights and the same phase~\cite{Casten1990}. Such coherence strongly enhances the observable transition strength. 
%an ideal coolective state: all the 2qp configurations would contribute equally with a statistical weight of $1/N_{\mathrm{2qp}}$. 

For a given excited state $|w\rangle$, the contribution of a specific 2qp configuration can be quantified using the QRPA amplitudes, 
\begin{equation}
 a^w_{2qp} \equiv (X^w_{2qp})^2 - (Y^w_{2qp})^2,
\end{equation}
which satisfy the normalization condition 
\begin{equation}\label{norm}
\left| \sum_{\mathrm{2qp=1}}^{N_{\mathrm{2qp}}} a^w_{\mathrm{2qp}} \right| = \left| \sum_{\mathrm{2qp=1}}^{N_{\mathrm{2qp}}} \left(X^2_\mathrm{{2qp}}(w) - Y^2_\mathrm{{2qp}}(w) \right) \right| = 1,
\end{equation}
%where $a^w_\mu \equiv (X^w_\mu)^2 - (Y^w_\mu)^2$ and
where $X$ and $Y$ represent forward and backward QRPA amplitudes, respectively.
Here $N_{\mathrm{2qp}}$ is the total number of possible 2qp excitations for a given $K^{\pi}$. 
%{\color{blue} (Sophie: I don’t see any normalization in the formula above, only a max value for the sum. The notion of normalisation is needed for eq 12. I would move this sentence after the eq 12.)}
This normalization ensures that the contribution of each 2qp configuration to the state $|w\rangle$ can be expressed as a percentage of the total. To distinguish the contributions from neutrons and protons, we separately sum their respective contributions, denoted as $\%(\nu)$ for neutrons and $\%(\pi)$ for protons.

We define the fragmentation number of a given state as counting the number of 2qp configurations whose contributions exceed the average weight $1/N_{\mathrm{2qp}}$ \cite{Colo2009, Martini2011, Vretenar2001}:
%$[(X_\mathrm{{2qp}}^2(w)-Y_\mathrm{{2qp}}^2(w)] \geq 1/N_{\mathrm{2qp}}$ \cite{Colo2009, Martini2011, Vretenar2001}.
%\begin{equation}
%N^* = \sum_{\mathrm{2qp}} \Theta \left( \left| X_{\mathrm{2qp}}^2(w) - Y_{\mathrm{2qp}}^2(w) \right| - \frac{1}{N_{\mathrm{2qp}}} \right),
%\end{equation}
\begin{equation}
 N^* = \sum_{\mathrm{2qp}} \Theta \left(a^w_{\mathrm{2qp}} - \frac{1}{N_{\mathrm{2qp}}} \right),
\end{equation} 
with the Heaviside step function $\Theta(x) = 1 $ for $ x \geq 0 $ and $\Theta(x) = 0 $ otherwise.
A large $N^*$ indicates high fragmentation, which is a necessary (though not sufficient) condition for collectivity. 
We also separately evaluate neutron ($N^*_\nu$) and proton ($N^*_\pi$) contributions.
The total number of 2qp configurations, $N_{\mathrm{2qp}}$, is determined by the size of the configuration space, which depends on the number of major oscillator shells included in the calculations. 
In this study, our calculations are performed in bases that span 11 major oscillator shells with a maximum oscillator shell number of 10, as detailed in the following section, resulting in $N_{\mathrm{2qp}}=12152$ for $K^{\pi}=0^{-}$. 
For spherical nuclei, the results for $K=0$ and $K=1$ are strictly equivalent. We perform the analysis in $K=0$ 
%to ensure a consistent configuration space 
for all nuclei considered.

We measure the degree of coherence by computing the energy shift $\delta E_w$, defined as the difference between the QRPA state energy $E_w$ and the weighted average of unperturbed 2qp energies $E_\mu$ ~\cite{Emanuel2019}: 
\begin{equation} 
	\delta E_w = E_w - \sum_\mu E_\mu \left| a^w_\mu \right|.
\end{equation}
The constructive contributions of individual 2qp configurations enhance coherence, which manifests as large energy shifts relative to the unperturbed 2qp energies.
%{\color{blue} The constructive contributions of individual 2qp configurations enhance coherence, lowering the energy of the state and thus resulting in negative values of the energy shift $\delta E$.}
By analyzing $\delta E$ together with the total $B(E1)$ strength in Weisskopf units (W.u.) for each state, we evaluate whether the collectivity inferred from the wave function is consistently manifested in observable transition strengths.

%Do not remove below:
%In nuclear physics, collectivity refers to the degree to which nucleons (protons and neutrons) in a nucleus participate in a cooperative or collective manner during an excitation or transition.
%In a highly collective state, many nucleons move in a correlated fashion, often leading to phenomena such as nuclear vibrations or rotations that involve large parts of the nucleus. For example, in a collective dipole state, protons and neutrons may oscillate together against each other, creating a strong dipole moment. In contrast, in a less collective or more single-particle-like state, only a few nucleons are involved in the excitation, and their motion is largely independent of the others.
%Collectivity is often measured by examining the number of two-quasiparticle (2qp) configurations that significantly contribute to an excited state. If a large number of 2qp configurations contribute, the state is considered more collective. Conversely, if only a few 2qp configurations contribute, the excitation is more single-particle-like and less collective.

\subsection{Numerical convergence and HFB ground state properties}

\begin{figure} [h!]
%\captionsetup[figure]{justification= raggedleft, singlelinecheck=false}
\centering
\includegraphics[width=0.95\textwidth]{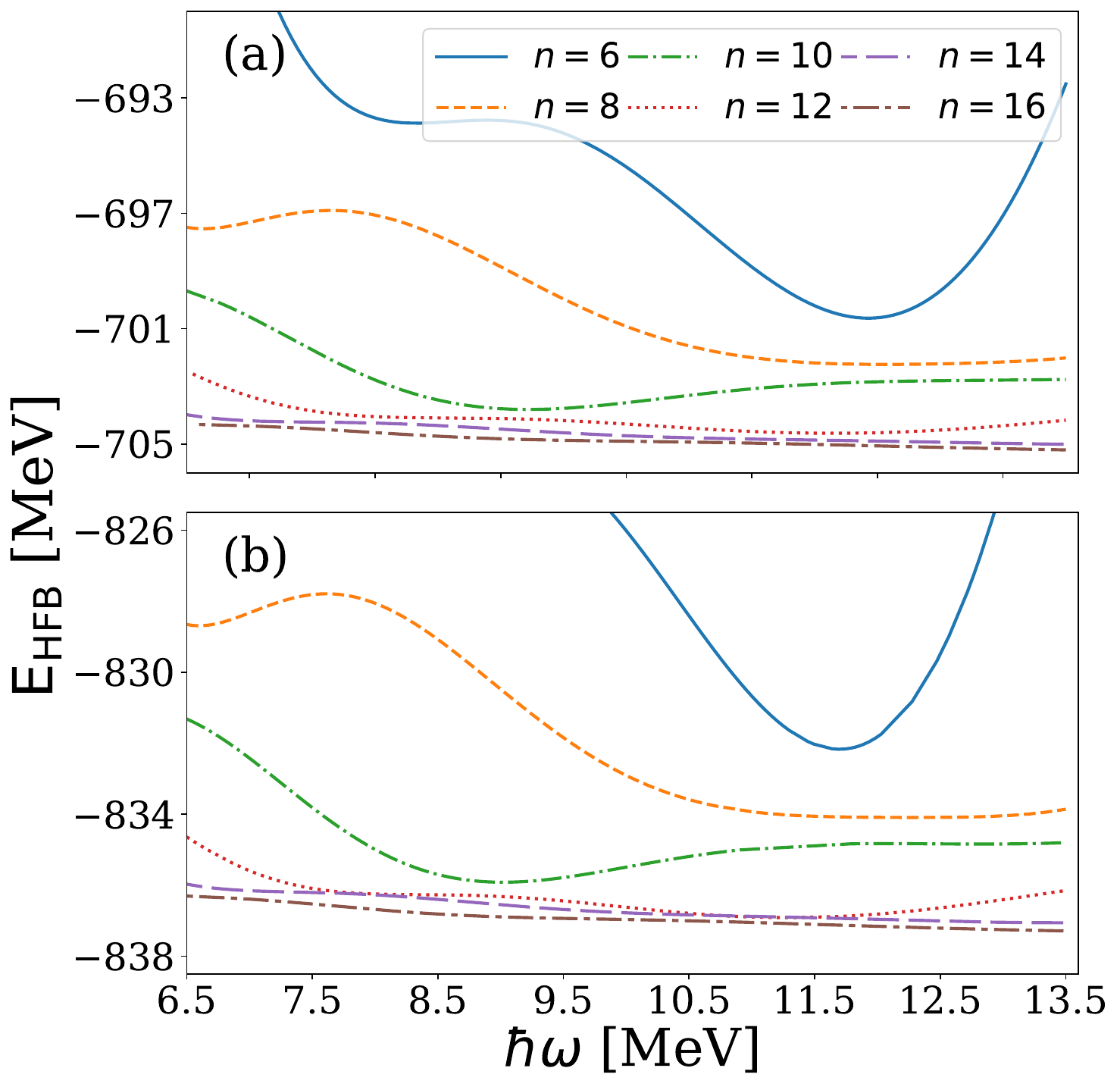}\vspace{0.001cm}
%\captionsetup{justification=raggedleft, singlelinecheck=false}
\caption{\label{Nosc} HFB energies as functions of the HO parameter $\hbar w$ for maximum shell number $n$ ranging from 6 to 16 for (a) $^{84}$Mo and (b) $^{96}$Mo.  
In this study, $n=10$ maximum shell number is chosen as the basis size to ensure accurate descriptions of Mo isotopes ($A=82$–$98$) while balancing solution stability and computational efficiency.
} 
\end{figure}

Reliable convergence within the mean-field formalism requires selecting an optimal number of major HO shells. While increasing the basis size enhances accuracy, it also increases computational demands, particularly for QRPA calculations. To balance precision and efficiency, we examine convergence behavior within the region of interest and determine an appropriate limit for the HO basis size.
Figure~\ref{Nosc} illustrates the HFB energy as a function of the HO parameter $\hbar w$ for both $^{84}$Mo and $^{96}$Mo. Our calculations are performed in bases with maximum shell number $n$ (i.e., they span $n+1$ major oscillator shells, starting at 0).
%for major shell numbers $n+1$ ranging from 6 to 16, where $n$ denotes the maximum shell number. Results are presented for both $^{84}$Mo and $^{96}$Mo.}
The values of $\hbar w$ are chosen to minimize the HFB energy for each $n$. As shown, $n=16$ closely approximates an infinite basis, with difference in the HFB energy below 1 MeV across $\hbar w$ values between 6.5 and 13.5.
Table~\ref{t-Nosc} summarizes the HFB energy (E$_{\mathrm{HFB}}$), charge radius ($R_{\text{ch}}$), and quadrupole deformation parameter ($\beta_2$) for various $n$. The results demonstrate stable HFB solutions as $n$ increases from 10 to 16, with the HFB energy for $n=10$ differing by less than $0.2 \,\%$ from $n=16$.
For this study, we adopt $n=10$, as the basis size, sufficient to accurately describe both proton- and neutron-rich molybdenum isotopes in the mass range $A=82$ to $A=98$. Additionally, $n=10$ provides a good compromise between stability of the solutions and computational cost of the QRPA calculations.
%{\color{blue} Which $\hbar w$ value?}

\begin{table}
%\captionsetup[table]{justification=raggedleft, singlelinecheck=false}
\caption{\label{t-Nosc} HFB energies (E$_{\mathrm{HFB}}$), charge radii ($R_{\text{ch}}$), and quadrupole deformation parameters ($\beta_2$) for $^{84}$Mo and $^{96}$Mo calculated using HO bases with varying shell numbers $n$. The results demonstrate stable values as $n$ increases, with acceptable variation between $n=10$ and $n=16$.}
\centering
\begin{tabular}{c |c | c| c| c| c| c|} 
\hline
Isotope & Property & $n=6$ & $n=8$ & $n=10$ & $n=12$ & $n=16$ \\
\hline 
\hline
\multirow{3}{*}{$^{84}$Mo} & E$_{\mathrm{HFB}}$ & -700.64 & -702.24 & -703.80 & -704.63 & -705.20 \\ 
& $R_{\text{ch}}$ & 4.25 & 4.25 & 4.26 & 4.26 & 4.26 \\ 
& $\beta_2$ & 0.0  & 0.0 & 0.0 & 0.0 & 0.0  \\ 
\hline
\multirow{3}{*}{$^{96}$Mo} & E$_{\mathrm{HFB}}$ & -832.17 & -834.09 & -835.91 & -836.91 & -837.28 \\ 
& $R_{\text{ch}}$ & 4.30 & 4.32 & 4.32 & 4.32 & 4.32 \\ 
& $\beta_2$ & 0.0  & 0.0 & 0.0 & 0.0 & 0.0 \\
\hline
\end{tabular}
\end{table}

%\clearpage
\section{Results}

\subsection{Neutron and proton spatial densities for spherical $^{82-98}$Mo isotopes}

The skin thickness, $\delta_{np}$, is defined as the difference between the root-mean-square (rms) radii of neutrons and protons, 
%\begin{equation}
%\delta_{np}=\sqrt{\left< r_n^2 \right>} - \sqrt{ \left< r_p^2 \right>},
%\end{equation}
\begin{equation}
\delta_{np} = \sqrt{\left< r_n^2 \right> \vphantom{\left< r_p^2 \right>}} - \sqrt{ \left< r_p^2 \right> }
\end{equation}
where $\left< r^2_q \right> = \frac{1}{A_q} \int d^3r \, r^2 \rho_q(\vec{r})$ denotes the mean square radius of the ground state density distribution $\rho_q$ for neutron ($q=n$) and proton ($q=p$), respectively. These distributions are normalized to the respective particle number $A_q=N,Z$. The calculated $\delta_{np}$ values are presented in Table~\ref{GS-skin}. A negative $\delta_{np}$ indicates a proton skin. With $\delta_{np} \sim 0$ in $^{90}$Mo, this isotope serves as a convenient reference. Relative to it, the neutron skin becomes increasingly pronounced in $^{92-98}$Mo, whereas the proton skin develops in $^{88-82}$Mo.

\begin{table}[h!]
\captionsetup[figure]{justification=justified, singlelinecheck=false}
\caption{\label{GS-skin} Neutron and proton rms radii ($R_n$, $R_p$) and the corresponding $\delta_{np}$ values (in fm) for $^{82-98}$Mo isotopes.}
%\begin{center}
%\begin{tabular}{|c |c |c |c ||c |c |c |c |} 
%\hline
%Isotopes & $R_n$ & $R_p$ & $\delta_{np}$ & Isotopes & $R_n$ & $R_p$ & $\delta_{np}$  \\\hline 
%$^{82}$Mo & 4.075 & 4.177 & -0.102 & $^{92}$Mo & 4.237 & 4.215 & 0.023 \\ \hline
%$^{84}$Mo & 4.111 & 4.184 & -0.073 & $^{94}$Mo & 4.281 & 4.232 & 0.049 \\ \hline
%$^{86}$Mo & 4.145 & 4.191 & -0.046 & $^{96}$Mo & 4.323 & 4.250 & 0.073 \\ \hline
%$^{88}$Mo & 4.177 & 4.198 & -0.021 & $^{98}$Mo & 4.365 & 4.269 & 0.096 \\ \hline
%$^{90}$Mo & 4.208 & 4.215 & 0.002 & & & & \\
%\hline
%\end{tabular}
%\end{center}
%\end{table}
\begin{center}
\begin{tabular}{|c |c |c |c |} 
\hline
Isotopes & $R_n$ (fm) & $R_p$ (fm) & $\delta_{np}$ (fm) \\ \hline 
$^{82}$Mo & 4.075 & 4.177 & -0.102 \\ \hline
$^{84}$Mo & 4.111 & 4.184 & -0.073 \\ \hline
$^{86}$Mo & 4.145 & 4.191 & -0.046 \\ \hline
$^{88}$Mo & 4.177 & 4.198 & -0.021 \\ \hline
$^{90}$Mo & 4.208 & 4.215 & 0.002 \\ \hline
$^{92}$Mo & 4.237 & 4.215 & 0.023 \\ \hline
$^{94}$Mo & 4.281 & 4.232 & 0.049 \\ \hline
$^{96}$Mo & 4.323 & 4.250 & 0.073 \\ \hline
$^{98}$Mo & 4.365 & 4.269 & 0.096 \\ \hline
\end{tabular}
\end{center}
\end{table}

\begin{figure} [h!]
\centering
\includegraphics[width=\textwidth]{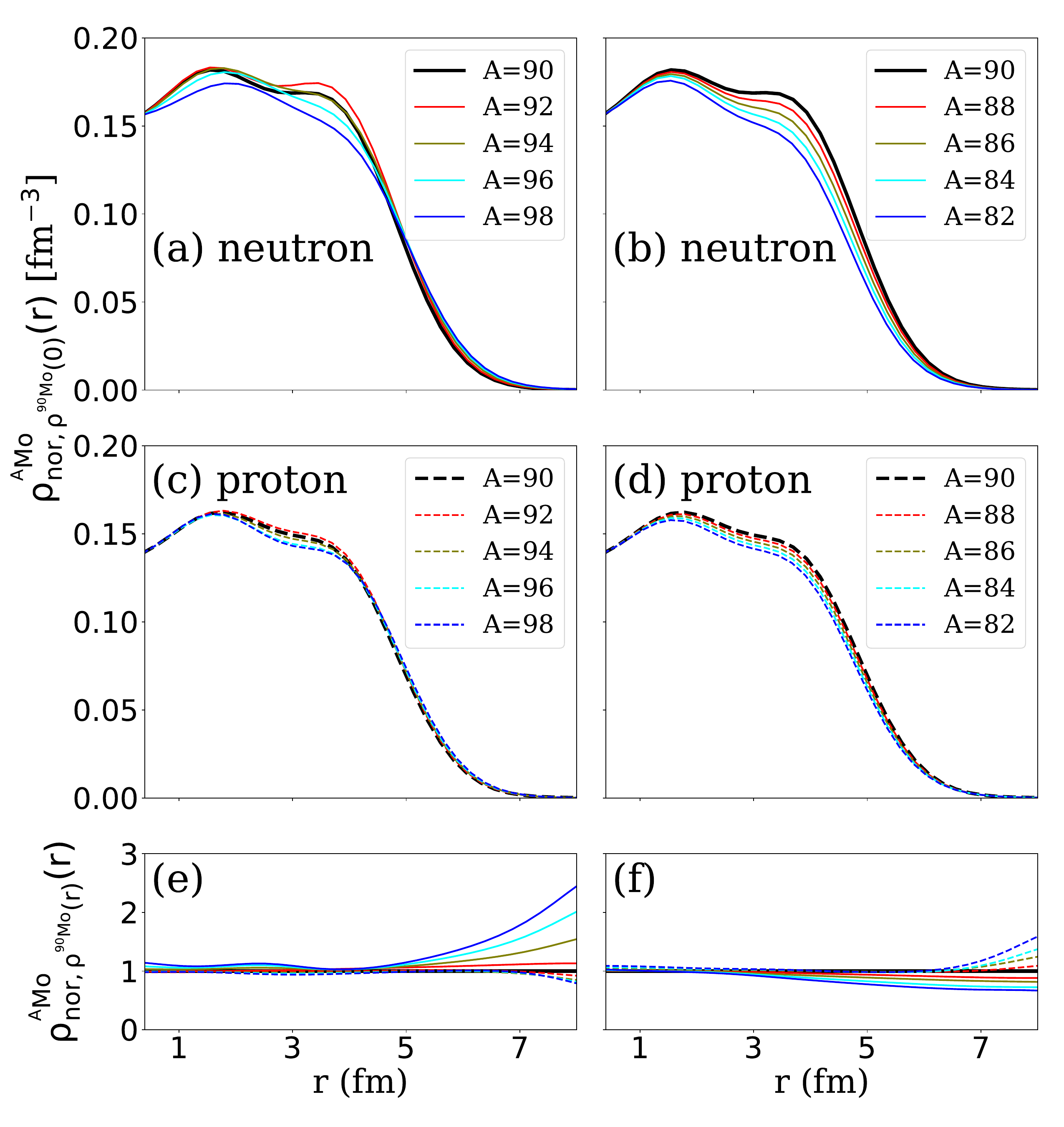}\vspace{0.01cm}
\caption{\label{GS-den} Ground state spatial densities of neutrons (solid lines) and protons (dashed lines) for $^{82-98}$Mo isotopes, calculated using the D1M Gogny interaction. Panels (a+c) and (b+d) show neutron and proton densities for $^{90-98}$Mo and $^{82-90}$Mo, respectively, normalized by $\rho_{\mathrm{^{90}Mo}}(0)/\rho_{\mathrm{^{A}Mo}}(0)$ to highlight the gradual outward extension relative to $^{90}$Mo. Panels (e) and (f) present densities normalized by $1/\rho_{\mathrm{^{90}Mo}}(r)$, illustrating the growth of neutron and proton skins above the surface region (5-8 fm) relative to $^{90}$Mo. 
}
\end{figure}

Figure~\ref{GS-den} presents the spatial densities of protons and neutrons in the ground states of spherical $^{82-98}$Mo isotopes, taking $^{90}$Mo as a reference nucleus.
To facilitate comparison, panels (a-d) display neutron ((a) and (b)) and proton ((c) and (d)) spatial densities normalized according to $\rho_q^{^A\mathrm{Mo}}(r) \times \rho_q^{^{90}\mathrm{Mo}}(0)/\rho_q^{^A\mathrm{Mo}}(0)$ to align the density values at $r=0$ with those of $^{90}$Mo. 
In Fig.~\ref{GS-den} (a), the neutron densities for $^{90-98}$Mo exhibit similar distributions within the nuclear interior ($r \leq 4.3$ fm). Beyond 5 fm, however, their tails extend increasingly outward and consistently exceed the $^{90}$Mo distribution, reflecting the addition of neutrons. 
In contrast, the proton densities in Fig.~\ref{GS-den} (c) remain nearly unchanged across these isotopes, indicating the development of a neutron skin. This trend is clearly seen in Fig.~\ref{GS-den} (e), where the densities normalized by $\rho_q^{^A\mathrm{Mo}}(r) \times 1/\rho_q^{\mathrm{^{90}Mo}}(r)$ reveal the gradual development of the neutron skin in the 5 to 8 fm region. 
For $^{82-90}$Mo isotopes, Figs.~\ref{GS-den} (b) and (d) show that both neutron and proton densities in $^{82-88}$Mo are consistently lower than those in $^{90}$Mo across the entire radial range. Figure~\ref{GS-den} (f) further illustrates the development of proton skins in these isotopes, as indicated by enhanced proton densities and reduced neutron densities in the 5-8 fm region relative to $^{90}$Mo.

%The densities are normalized by the factor $\rho_{\mathrm{^{90}Mo}}(0)/\rho_{\mathrm{^{A}Mo}}(0)$ to highlight the presence of neutron and proton skins in the surface region, approximately at and beyond 4 fm. For isotopes with A$\,\geq\,$90, the neutron density extends to larger radial distance than the proton density, up to $^{98}$Mo, indicating the formation of a neutron skin in $^{90-98}$Mo isotopes. Conversely, a proton skin is observed in $^{82-88}$Mo isotopes. 

\subsection{Dipole response near the $S_n$ and its correlation with skin thickness}

\begin{figure}
\centering
\includegraphics[width=0.95\textwidth]{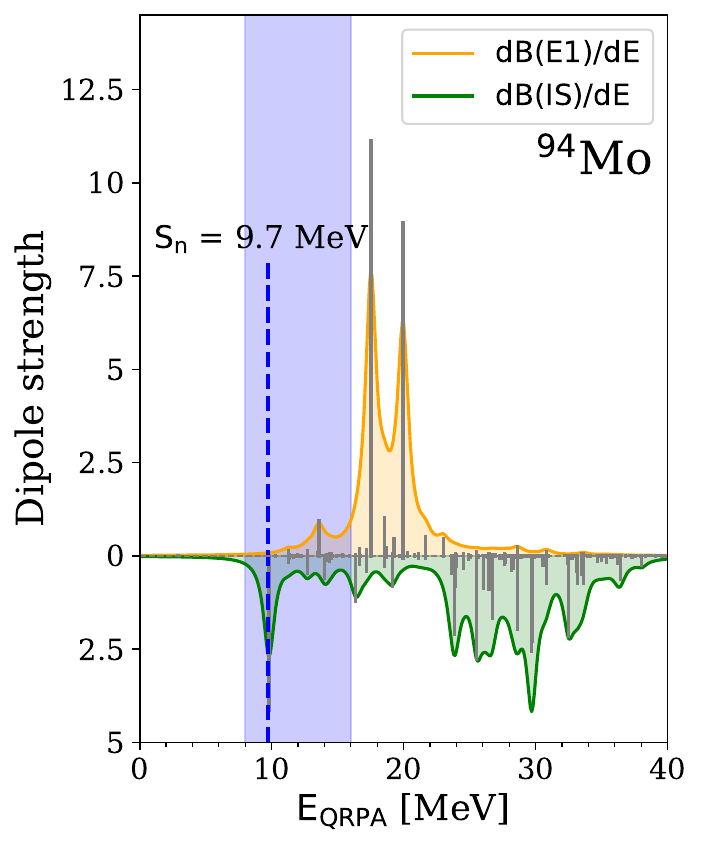}
\vspace{0.01cm}
\caption{\label{BE1-80-98} Electric ($B(E1)$ [e$^2$fm$^2$]) and isoscalar ($B(IS)$ [10$^{3}$e$^2$fm$^6$]) dipole strengths as a function of the QRPA excitation energy for $^{94}$Mo. The curves ($dB(E1)/dE$ [e$^2$fm$^2$MeV$^{-1}$] and $dB(IS)/dE$ [10$^{3}$e$^2$fm$^6$MeV$^{-1}$]) are obtained by smoothing the discrete spectra (gray lines) with a Lorentzian of 1 MeV width. The $S_n$ threshold is indicated by the blue dashed line. The blue shaded region highlights the enhancement region discussed in this work.
}
\end{figure}

% Do not remove below sentences:
%The definition of the PDR varies across different studies [Refs]. In some cases, it is regarded as the low-energy component of the dipole strength distribution [Refs], while in others, it is described as a macroscopic bell mode [Refs] or an isoscalar (IS) dipole resonance [Refs]. Given these variations, it is essential to establish clear criteria for identifying PDR candidates in our study. We define PDR candidates based on their radial transition density characteristics, which provide insight into the underlying nuclear structure by revealing how neutron and proton densities oscillate during excitation.
%The definition of the PDR varies across different studies [Refs]. In some cases, it is regarded as the low-energy component of the dipole strength distribution [Refs], while in others, it is described as a macroscopic bell mode [Refs] or an isoscalar (IS) dipole resonance [Refs]. 

Figure~\ref{BE1-80-98} illustrates the $E1$ strength ($B(E1)$ in Eq.~\eqref{BIV}) and $IS$ dipole strength ($B(IS)$ in Eq.~\eqref{BIS}) distributions for $^{94}$Mo. The discrete spectra are smoothed using a Lorentzian function of width $\Gamma = 1$ MeV. 
A small but distinct enhancement in the $E1$ strength appears near the $S_n$ threshold, as marked by the blue shaded region. Similar features observed in experimental scattering data have been interpreted as signatures of low-energy dipole excitations, the so-called PDR \cite{Bracco2019}. To further shed light on this small enhancement phenomenon, we examine the structure of the calculated dipole states near the $S_n$ threshold.

To examine the enhanced $E1$ strength near the $S_n$ threshold, we divide the $E1$ strength into two regions: the region containing the enhanced low-energy dipole structure, separated from the major GDR peak and often referred to as the PDR region, and the GDR region at higher excitation energies.
This separation is guided by the fact that the PDR region typically accounts for only a small fraction of the energy-weighted sum rule (EWSR). The EWSR is defined as the first energy-weighted moment of the $E1$ strength, $m_1 = \sum_w E_w B_w(E1)$, where $E_w$ and $B_w(E1)$ are the QRPA excitation energy and transition strength of the $w$-th QRPA state, respectively.
The fraction of the EWSR exhausted below a given cutoff energy $E_{\text{cut}}$ is defined as $m_1^{(E_w < E_{\text{cut}})} / m_1^{\text{total}}$. By evaluating this quantity across several cutoff values, we find that the $^{82-98}$Mo isotopes exhaust approximately 3–6 $\%$ of the EWSR below 16 MeV, 3–8 $\%$ below 17 MeV, and 40–50 $\%$ below 18 MeV. Based on these results, we adopt 16 MeV as the upper boundary of the enhanced low-energy dipole region throughout this work.

%Do not remove: criteria for dividing energy region
%a clustering of dipole states close to the $S_n$
%Here, the PDR mode has to be distinguished from the other known low-energy isoscalar dipole excitations, namely, the two-phonon $1^-$ states resulting from the anharmonic interactions of the lowest $2^+$ and $3^-1$ states in a nucleus.
%It turns outthat while the effect on the monopole and quadrupole states is moderate, the one on the dipole states is stronger [I.Hamamoto, PRC 53 (1996) 765, I.Hamamoto,PRC 53 (1996) R1492, F.Catara, NPA 614 (1997) 86, F.Catara, NPA 624 (1997) 449, I.Hamamoto, PRC 57 (1998) R1064.]. 

\begin{figure} [h!]
\centering
\includegraphics[width=0.99\textwidth]{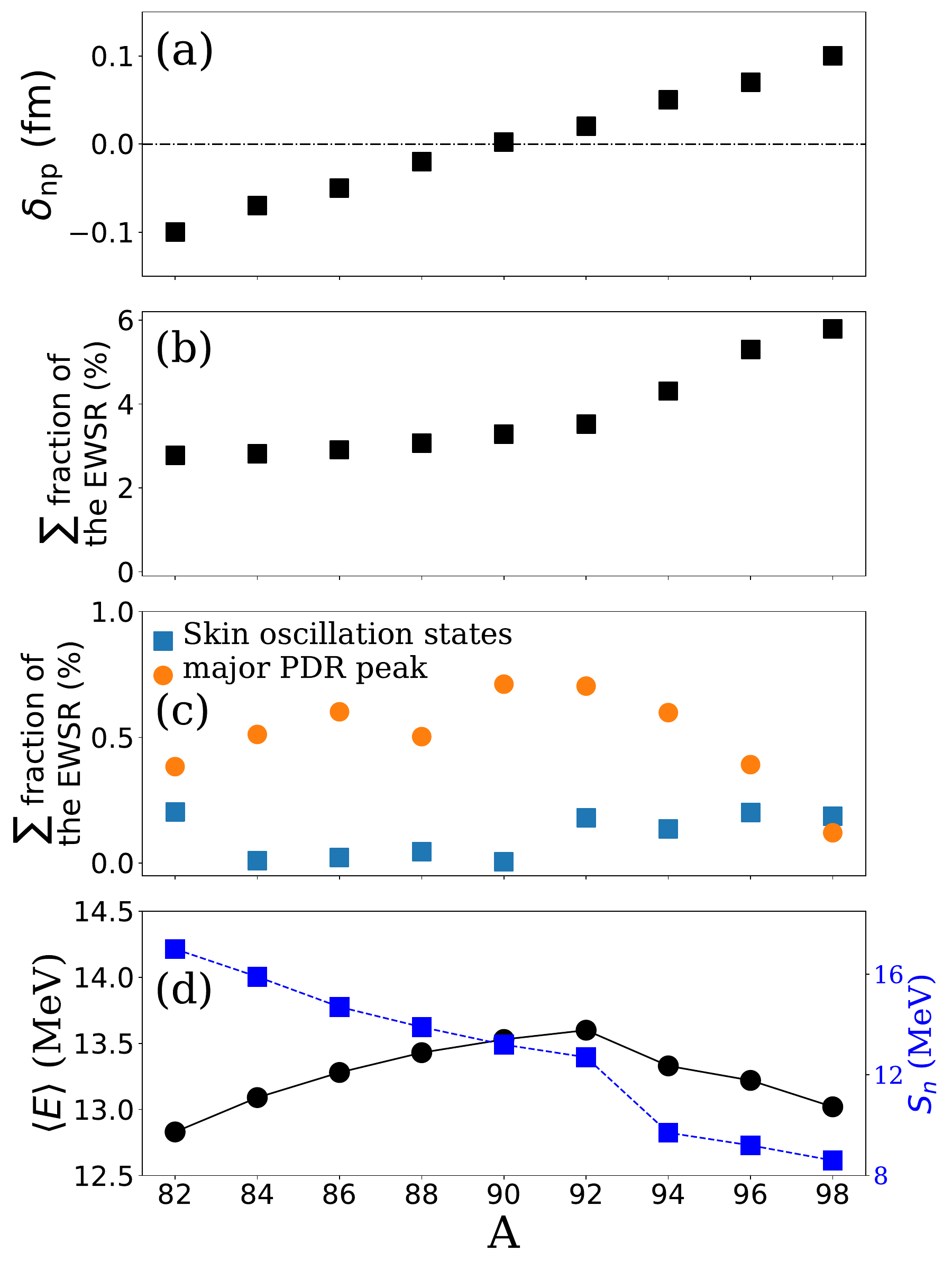}\vspace{0.01cm}
\caption{\label{EWSR} Correlation between ground-state skin thickness and low-energy dipole strength in spherical Mo isotopes, $^{82-98}$Mo. Negative values indicate a proton skin. The top panel (a) shows the neutron skin thickness $\delta_{np}$. Panel (b) presents the summed $E1$ strength in the low-energy dipole region (8-16 MeV) from all QRPA dipole states, expressed as a fraction of the total EWSR up to 40 MeV. Panel (c) shows the summed $E1$ strength in the low-energy region (8-16 MeV), where the contributions from skin oscillation states (blue squares) and major PDR peaks (orange circles) are separately shown, each expressed as a fraction of the total EWSR in the 8-16 MeV window. 
The bottom panel (d) displays the average excitation energies $\left\langle E \right\rangle$ of all QRPA dipole states in the 8-16 MeV region (black solid line), compared with the $S_n$ threshold (blue dashed line). 
}
\end{figure} 

It is of particular interest to explore the connection between the enhanced low-energy dipole modes and the development of neutron and proton skins.
%An interesting question is how the low-energy dipole modes are connected to the development of neutron and proton skins. 
Figure~\ref{EWSR} displays the correlation between the enhanced low-energy dipole response and the ground-state skin thickness. In panel (a), we plot the ground-state neutron skin thickness $\delta_{np}$. 
Panel (b) presents the summed $E1$ strength from QRPA dipole states in the 8–16 MeV window, expressed as a fraction of the total EWSR up to 40 MeV.
%Panel (b) presents the summed EWSR fraction exhausted by all QRPA dipole states in the 8-16 MeV window. 
%Three sets of data are shown: the total contribution from all QRPA dipole states (black squares), the subset of states selected based on their transition density features (red circles), and the major PDR peak (olive triangles), those expressed as percentages. The selection criteria for the red circles will be discussed in detail later.
For $^{90-98}$Mo, the low-energy dipole strength increases smoothly with neutron number, in parallel with the development of a neutron skin. This trend supports previous theoretical studies that the enhancement may be correlated with surface neutron excess \cite{Tsoneva2008, Tsoneva2004}.
In contrast, the $^{82-88}$Mo nuclei, which are closer to the $N=Z$ limit, exhibit a proton skin despite some of them having more neutrons than protons, likely due to the interplay of Coulomb repulsion and strong nuclear forces. In this region, the EWSR fraction remains nearly constant, indicating the low-energy dipole strength is relatively less sensitive to changes in skin thickness.
%This suggests that, in nuclei near the $N=Z$ limit, where the proton skin persists due to the interplay of Coulomb repulsion and strong nuclear forces, the low-energy dipole strength is less sensitive to changes in skin thickness.
%A sudden drop in the summed EWSR fraction is observed for $^{98}$Mo. This may be associated with the onset of deformation in heavier Mo isotopes, particularly $^{100}$Mo, which appears deformed in our calculations. A more detailed analysis, including deformation effects, will be pursued in future work.

In recent literature, the location of the PDR with respect to the $S_n$ threshold has been actively discussed. Experimental studies using ($\gamma, xn$) reactions have measured the enhanced $E1$ strength both below and above the $S_n$ threshold~\cite{Rusev2006,Rusev2009,Erhard2010}. 
From a theoretical perspective, Ref.~\cite{Paar2005PLB-energy} examined the evolution of the PDR by tracking the energy of the most prominent peak, and found that its location shifts relative to the $S_n$ threshold across isotopes. 
In our analysis, we consider the average excitation energy of the low-energy dipole states, %Fig.~\ref{EWSR} (c) shows that the average excitation energies of the low-energy dipole states,
calculated as $\left\langle E \right\rangle=\sum_w E_w\,B_w(E1) / \sum_w B_w(E1)$ (in MeV), where the sum runs over all dipole states in the 8-16 MeV energy window. As shown in Fig.~\ref{EWSR} (d), these average energies tend to decrease with the development of either neutron or proton skins. A transition is seen around $A=90$: for $^{82-88}$Mo, the average excitation energies lie below the $S_n$ threshold, whereas for $^{90-98}$Mo, they appear above the $S_n$ threshold. 
%In the neutron-rich region ($A \geq 90$), 
This behavior is accompanied by a modest discontinuity near the shell closure at $A=92$, where a sharp drop in $S_n$ is observed. This pattern is similar to the trends reported in Ni, Sn, and Pb isotopes~\cite{Paar2005PLB-energy}, though the specific definitions of the PDR energy differ.
Figure~\ref{EWSR} (c) will be discussed in Section~\ref{Sec-nature} below.
%Regardless of the energy location of these dipole states relative to $S_n$, we note that the nature of the excitation—whether it constitutes a mode distinct from the GDR—can only be properly established through the analysis of transition densities, as discussed in the following section.
%Nevertheless, whatever is the location of the low-energy dipole state, we note that the only way to establish its nature, namely, that it is a different mode with respect to the GDR one, is to look at the transition densities, as discussed in the following section.

%\begin{table}
%\caption{\label{aveE} Average energies calculated using $\left\langle E \right\rangle=\sum_i E_i\,B_i / \sum_i B_i$ (in MeV), where $E_i$ and $B_i$ are the QRPA energies and reduced transition probabilities, respectively. Each dipole states are in the energy region of 8 to 16 MeV.}
%\begin{center}
%\begin{tabular}{|c |c |c |c |c |c |} 
%\hline
%Isotopes & $S_p$ & $S_n$ & PDR & GDR & total  \\ \hline\hline 
%$^{82}$Mo & 1.3 & 17.0 & 12.8 & 19.2 & 19.0   \\ \hline
%$^{84}$Mo & 3.8 & 15.9 & 13.1 & 19.3 & 19.0 \\ \hline
%$^{86}$Mo & 5.1 & 14.7 & 13.3 & 19.3 & 19.1 \\ \hline
%$^{88}$Mo & 6.1 & 13.9 & 13.5 & 19.4 & 19.1 \\ \hline
%$^{90}$Mo & 6.8 & 13.2 & 13.6 & 19.4 & 19.2 \\ \hline
%$^{92}$Mo & 7.5 & 12.7 & 13.6 & 19.5 & 19.2 \\ \hline
%$^{94}$Mo & 8.5 & 9.7 & 13.4 & 19.3 & 19.0 \\ \hline
%$^{96}$Mo & 9.3 & 9.2 & 13.3 & 19.1 & 18.7  \\ \hline
%$^{98}$Mo & 9.8 & 8.6 & 13.0 & 19.0 & 18.6 \\ 
%\hline
%\end{tabular}
%\end{center}
%\end{table}

\subsection{Nature of low-energy dipole states}\label{Sec-nature}

\begin{figure*} 
\centering
\includegraphics[width=0.99\textwidth]{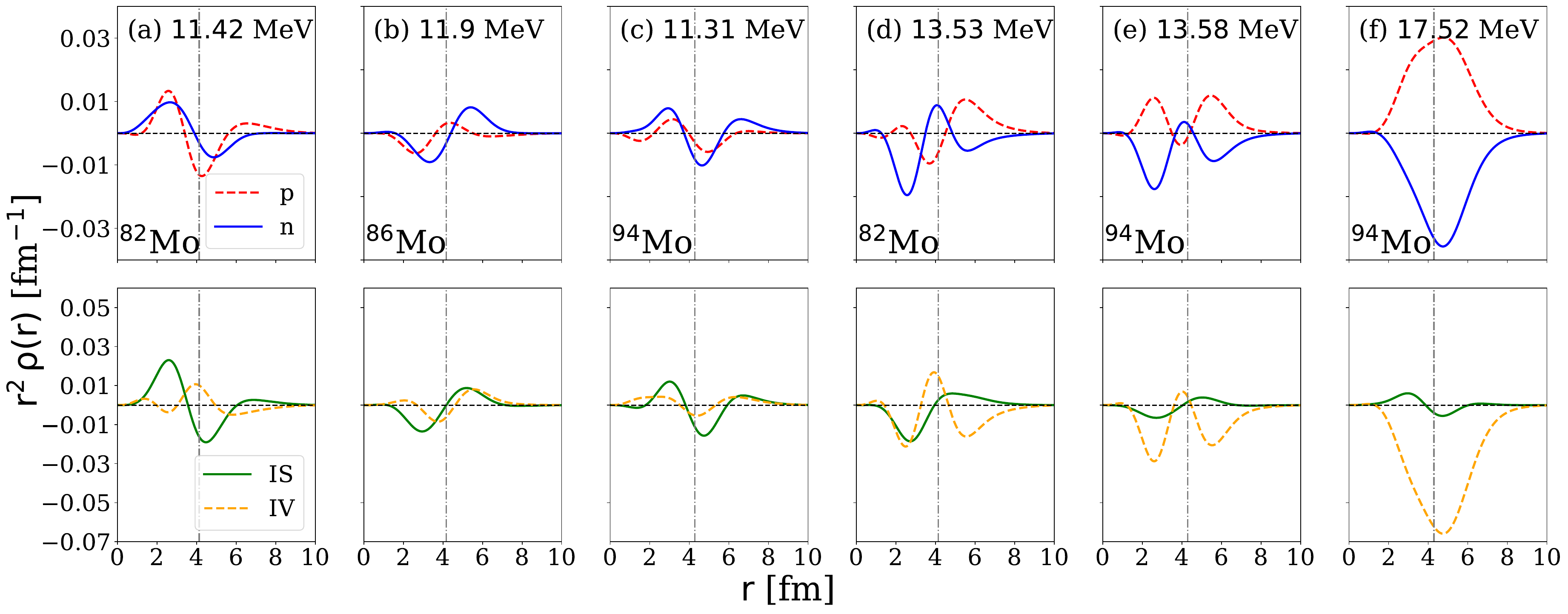}%\hspace{0.01cm}
\caption{\label{radial_td} Radial transition densities for representative dipole states in Mo isotopes: (a) a low-energy state at 11.42 MeV in $^{82}$Mo, (b) a low-energy state at 11.9 MeV in $^{86}$Mo, (c) a low-energy state at 11.31 MeV in $^{94}$Mo, (d) the major peak with the largest $B(E1)$ value within the enhanced low-energy region (major PDR peak) at 13.53 MeV in $^{82}$Mo, (e) the corresponding major PDR peak at 13.58 MeV in $^{94}$Mo, and (f) the major GDR peak exhibiting the largest $B(E1)$ value at 17.52 MeV in $^{94}$Mo. 
In each panel, the top subfigure shows the proton (red dashed line) and neutron (blue solid line) transition densities. The vertical grey dashed lines indicate the ground-state rms radii from HFB calculations. The bottom subfigure present the corresponding $IS$ (green solid line) and $IV$ (yellow dashed line) transition densities, highlighting the distinct features between low-energy and giant dipole states.
}
\end{figure*}

To investigate the underlying nature of dipole states in the low-energy enhancement region, Fig.~\ref{radial_td} displays the proton, neutron, $IS$, and $IV$ transition densities for representative dipole states: (a) a low-energy state at 11.42 MeV in $^{82}$Mo, (b) a low-energy state at 11.9 MeV in $^{86}$Mo, (c) a low-energy state at 11.31 MeV in $^{94}$Mo, (d) the major peak with the largest $B(E1)$ value within the low-energy enhancement window (major PDR peak) in $^{82}$Mo, (e) the corresponding major PDR peak   
%the major peak {\color{blue} with the largest $B(E1)$ value} within the low-energy enhancement window {\color{blue} (major PDR peak)} 
in $^{94}$Mo, and (f) the major GDR peak with the largest $B(E1)$ value in $^{94}$Mo.
In $^{94}$Mo, the states in panels (e) and (f) of Fig.~\ref{radial_td} correspond to 
%the major low-energy and GDR peaks, respectively, as
those indicated in Fig.~\ref{BE1-80-98}.
As shown in panels (a)-(c), within the nuclear interior ($r \lesssim 4.3$ fm, corresponding to the rms ground-state radii) and surface region, the proton and neutron densities oscillate in phase, indicating a dominant $IS$ nature. 
Beyond the surface ($r \geq 6$ fm), the transition densities are dominated by one nucleon species (either neutrons or protons) while the other is nearly inactive. In this region, the $IS$ and $IV$ transition densities appear nearly identical in magnitude. 
The behavior observed in these low-energy states is consistent with previous theoretical findings by several authors \cite{Vretenar2001, Tsoneva2008, Colo2009}. In addition, the dominant nucleon species typically corresponds to the type of the ground-state skin, with one notable exception in panel (b): although $^{86}$Mo exhibits a proton skin but neutron excess, the 11.9 MeV state shows neutron-dominated oscillation in the external region.
Interestingly, in $^{90}$Mo, where the neutron skin thickness $\delta_{np}$ is close to zero, two low-energy states at $E_{\mathrm{qrpa}}=$ 10.17 and 11.24 MeV, with relatively small $B(E1)$ values on the order of $10^{-4} \, e^2 \mathrm{fm}^2 \mathrm{MeV}^{-1}$, display neutron oscillations in the external region. 
These states are dominated by nearly balanced proton and neutron pair contributions by the 2qp configurations corresponding to $(2p_{1/2},3s_{1/2})_{\pi,\nu}$, $(2p_{1/2},2d_{3/2})_{\pi,\nu}$ and $(2p_{3/2},2d_{5/2})_{\pi,\nu}$ configurations
(see Table II in the supplementary material for further details on these configurations).
Fig.~\ref{radial_td} (f) shows the transition densities of the major GDR state in $^{94}$Mo. We find the characteristics of an $IV$-dominated collective mode: proton and neutron transition densities oscillate out of phase across all radii, and the $IV$ component dominates over the $IS$ component.
%We can observe oscillatory features of the $IS$ transition density, which lead to a vanishing $B(E1)$ due to cancellation in the matrix element.
%Although oscillations are present in the $IS$ transition density, they cancel in the transition matrix element, resulting in a vanishing $B(E1)$ value.
%In addition, the $IS$ transition density shows some oscillations, which leads to a zero value for the corresponding $B(E1)$.
These findings highlight the fact that the low-energy dipole states represent a mode distinct from the $IV$ GDR.
Unlike the strong $IV$ GDR mode, the low-energy states are dominated by $IS$, and eventually mixed isospin nature. This indicates that these low-energy dipole states are not simply the low-energy tail of the GDR, but instead represent distinct excitation modes.

While the observed enhancement in dipole strength near the $S_n$ threshold has been widely attributed to the PDR and interpreted as originating from skin oscillations, a closer inspection of the individual dipole states by transition densities reveals a more nuanced interpretation.
Figure~\ref{radial_td} (d) and (e) display the major peaks within the enhanced low-energy region in $^{82}$Mo and $^{94}$Mo, respectively. Their pattern resembles the one of the $IV$ GDR, but with some differences: protons and neutrons oscillate out of phase beyond the surface region, while their interior regions show either a stronger $IV$ or a nearly similar magnitude of $IS$ and $IV$. In addition, their pattern shows a flip between neutrons and protons at the surface region.

We also quantitatively assess the contribution of the major PDR peak and skin oscillation states to the low-energy enhancement. 
Figure~\ref{EWSR} (c) shows the fraction of the EWSR within the 8–16 MeV region, separately for the skin oscillation states identified by radial transition density analysis (blue squares) and the major PDR peaks (orange circles), normalized to the total EWSR strength in the same energy window.
The results show that the major PDR peak dominates the EWSR fraction in the low-energy enhancement region across all isotopes.
However, the contribution from the skin oscillation states depends on the type of skin that develops.
For isotopes from $^{84}$Mo ($N=Z$) to $^{90}$Mo ($\delta_{np} \sim 0$), although they have excess neutrons, a proton skin develops, and the contribution from skin oscillation states remains nearly negligible. 
By selectively removing the major PDR peak, the remaining low-energy enhancement is found to be almost entirely suppressed in this mass range, even though these isotopes exhibit skin oscillation states in the enhancement.
In contrast, for $^{92-98}$Mo, where a neutron skin becomes prominent, the relative contribution from skin oscillation states gradually increases, indicating that the development of a neutron skin enhances the role of skin oscillations in the low-energy dipole response. Similarly, in proton-rich nuclei such as $^{82}$Mo, where a proton skin develops, a significant contribution from proton skin oscillation states is observed.
These findings suggest that the need for a careful assignment of the PDR, taking into account both its macroscopic interpretation and microscopic structure, to accurately disentangle different dipole excitation modes.
Moreover, the nature of these excitations and whether they constitute modes distinct from the GDR can only be properly established through a detailed analysis of their transition density patterns. 

\subsection{Collectivity of dipole states}

\begin{figure} [h!]
\centering
\includegraphics[width=0.95\textwidth]{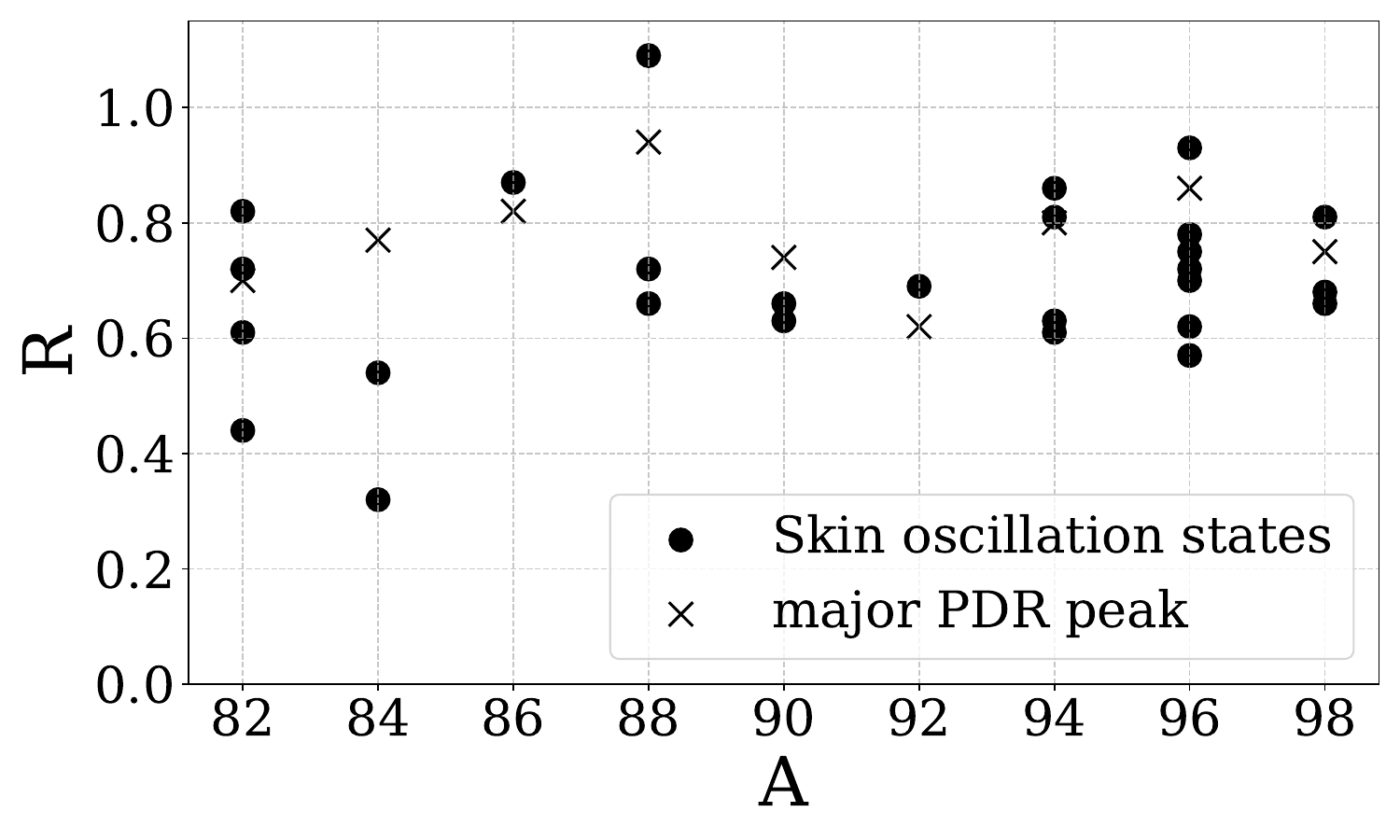}\vspace{0.01cm}
\caption{\label{R-A} Relative fragmentation ratio $R=N^*_{\text{P}}/N^*_{\text{G}}$ for skin oscillation states and major PDR peaks in $^{82-98}$Mo isotopes. }
\end{figure}

The collective nature of the PDR remains a subject of ongoing debate. Within the QRPA framework, collectivity is associated with the coherent superposition of multiple 2qp configurations in the excited state wave function.
To investigate this aspect, we introduce a relative fragmentation ratio $R=N^*_{\text{P}}/N^*_{\text{G}}$. Here $N^*_{\text{P}}$ represents the fragmentation number of either the skin oscillation or the major PDR peak, while $N^*_{\text{G}}$ is the fragmentation number of the major GDR peak, for each Mo isotope.
As shown in Fig.~\ref{R-A}, the majority of skin oscillation states in $^{82-98}$Mo exhibit $R$ values above 0.6. However, a few proton-dominated states in $^{82}$Mo and $^{84}$Mo display notably lower $R$ values between 0.3 and 0.5. Specifically the states at E=16.03 MeV in $^{82}$Mo and 16.63 MeV in $^{84}$Mo are dominated by a single proton configuration, $(1g_{9/2},2f_{7/2})_\pi$, contributing 92.1 $\%$ and 86.3 $\%$ of total amplitude, respectively, 
as detailed in Table I provided in the supplementary material.
These configurations involve weakly bound proton orbitals near the Fermi surface, indicating a possibility of single-particle-like nature. 
For context, the $R$ value is obtained relative to the major GDR peak, which serves as a reference for each isotope. In the case of $^{88}$Mo, this reference GDR state itself shows relatively low fragmentation compared to other isotopes, resulting in an $R$ value of 1.06 for the 12.01 MeV state.

\begin{figure} [h!]
\centering
\includegraphics[width=0.99\textwidth]{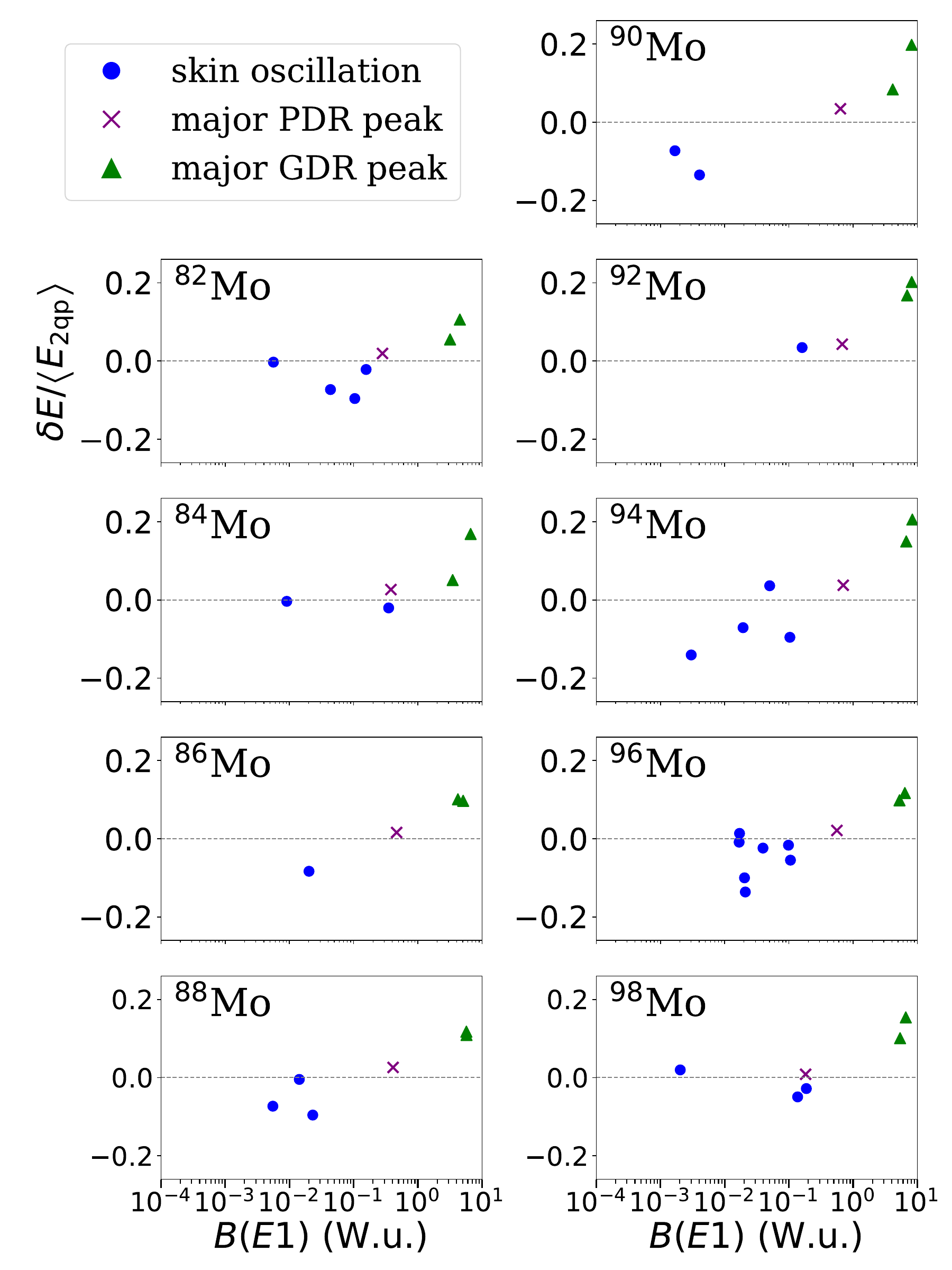}\vspace{0.01cm}
\caption{\label{Mo_Wu_deltaE} $B(E1)$ strength in Weisskopf units on the x-axis and relative energy shifts, $\delta E/\left< E_{\mathrm{2qp}}\right>$, on the y-axis for skin oscillation states, major PDR peaks, and major GDR peaks in $^{82-98}$Mo isotopes. Here, $\left< E_{\mathrm{2qp}}\right>$ denotes the average unperturbed 2qp energies.
}
\end{figure}

To further assess the collectivity of dipole states, we examine the coherence among their 2qp components and its relation to the observable transition strength. This is achieved by analyzing the correlation between the dipole strength $B(E1)$ and the relative energy shift $\delta E/ \left< E_{\mathrm{2qp}} \right>$, where $\left< E_{\mathrm{2qp}} \right>$ denotes the weighted average of unperturbed 2qp energies. 
Figure~\ref{Mo_Wu_deltaE} presents the results for $^{82-98}$Mo, with each panel showing the skin oscillation states, the major PDR peaks, and the major GDR peaks.
In the case of the major GDR peaks, the strong $IV$ nature leads to configurations that contribute constructively to the dipole operator with opposite phases between proton and neutron components. This results correlates with an upward energy shift and large $B(E1)$ values, a well-known signature of collective excitations.
In contrast, skin oscillation states in the PDR region exhibit a mixed nature of $IS$ and $IV$ components, typically dominated by the $IS$ contribution. Although many 2qp configurations participate, the $IS$ nature tends to suppress the electric dipole transition matrix element, often leading to smaller $B(E1)$ values. Ideally, coherent contributions would result in a large energy shift in magnitude. In practice however, phase differences among components can occur, suppressing the expected energy shift. 
These effects reflect the nuanced nature of collectivity in skin oscillation states: in some cases, despite sizable internal fragmentation, destructive interference can limit both the energy shift and the transition strength 
(see the discussion of Eqs.~\eqref{IVoperator} and ~\eqref{ISoperator}).
The major PDR peaks exhibit larger $B(E1)$ values compared to the skin oscillation states, but their relative energy shifts are typically smaller, suggesting a different type of configuration mixing and a weaker coherence among contributing components.

\section{Summary and Conclusion}

Electric dipole excitations in $^{82-98}$Mo isotopes were investigated using a fully consistent HFB+QRPA framework with the Gogny D1M finite-range effective interaction. Our calculations reveal the small enhancement of dipole strength near the $S_n$ threshold, a feature commonly associated with the appearance of PDR.
The states contributing to this enhancement exhibit distinct features. Analysis of their proton and neutron radial transition densities reveals that, within the nuclear interior, the two components oscillate in phase, indicating a dominant $IS$ character. At the surface and beyond, the transition densities are dominated by neutron components in $^{90-98}$Mo and by proton components in $^{82-88}$Mo, consistent with the development of neutron or proton skins, respectively. We also examined the major peaks within the PDR region. While the transition density magnitudes of the major peaks in the PDR region are comparable to those of the skin oscillation states, their patterns more closely resemble the GDR mode, though with noticeable differences in shape. The structure and shape of the transition densities clearly distinguish these low-lying states from the typical GDR pattern, with the GDR states exhibiting larger transition amplitudes.
Their mixed $IS$ and $IV$ character, often dominated by the $IS$ component, can cause partial cancellations in the electric dipole matrix elements, resulting in reduced $B(E1)$ strengths and reflecting the complex nature of their underlying structure.
%{\color{blue} Sophie: B(E1= electric dipole) is assumed to be due only to protons, but in the referential CdM. This sentence could be misinterpreted. Maybe remind the definitions 5 and 6 to explain how a large B(IS) can lead to a small B(E1)}

To contribute to the ongoing discussion regarding the collective nature of skin oscillation modes, we performed a detailed analysis of their internal configuration mixing and observable signatures. Our results based on the fragmentation ratio $R$ indicate that, while these modes exhibit lower fragmentation than the GDR, they cannot be described by a single 2qp configuration.
%In principle, ideal collective states arise when residual interactions cause significant deviations of the excitation energy from the unperturbed 2qp configurations, enabling the coherent superposition of many components. 
In principle, ideal collective states arise from strongly coherently mixed components of the unperturbed 2qp configurations. This in turn often increases the response of operators and can cause significant deviations of the excitation energy when compared to the unperturbed spectrum.
%Such coherence typically manifests as large transition strengths and substantial energy shifts relative to the unperturbed spectrum.
In practice, however, the situation is more subtle. Through the analysis of both the energy shifts $\delta E$ and the $B(E1)$ strengths, we observed that the GDR states, dominated by an $IV$ nature, exhibit constructive contributions that lead to large $B(E1)$ values and upward energy shifts. 
In contrast, skin oscillation states, typically dominated by an $IS$ nature, tend to suppress the electric dipole matrix element, resulting in comparatively smaller $B(E1)$ values. Although constructive coherence among 2qp components still exists, practical phase differences among components in the wave function limit the energy shift. 
Such features, though less pronounced than in the GDR, are observed in several skin oscillation states within the PDR region. The observed patterns reflect nontrivial coherent coupling among multiple 2qp configurations, where partial cancellations due to destructive interference can suppress the energy shifts and transition strengths, even when internal fragmentation is significant.
%These observations highlight that the suppression of transition strengths and energy shifts originates from different aspects of the underlying structure: the suppression of the dipole matrix elements due to the $IS$ nature, and the limited coherence among 2qp contributions.

We conclude that the feature discussed here indicate a promising way to study collectivity of new excitation modes. Further details on the microscopic two-quasiparticle configurations and transition density patterns are provided in the supplementary material. It would be worthwhile to extend the investigations into other mass regions as well, including deformed nuclei, where nuclear shape effects may significantly influence the excitation dynamics and isospin mixing of low-lying dipole modes. This may elucidate the complex relationship between nuclear deformation effects, skin development, and dipole strength fragmentation, ultimately contributing to a more complete picture across the nuclear chart.

%In particular, the states near 16 MeV in $^{82}$Mo and $^{84}$Mo are found to be dominated by nearly pure proton excitations involving weakly bound orbitals near the Fermi surface. These represent proton-driven skin oscillation and are noteworthy given that they occur in nuclei with $N$ slightly larger or equal to $Z$. Across the isotopic chain, most skin oscillation states are dominated by proton contributions, except for those around 12 MeV, where stronger neutron contributions appear. Interestingly, the trend of nearly balanced proton and neutron contributions observed in the 10$-$11 MeV states of $^{90}$Mo, where $\delta_{np}$ is approximately zero, continues in heavier isotopes such as $^{94,96,98}$Mo.

%Do not remove below pharagraphs:
%However, in nuclei with excess neutrons close to the $N=Z$ limit, where a proton skin persists due to the interplay of Coulomb repulsion and strong nuclear forces, the enhancement remains relatively unchanged. The analysis of transition densities revealed distinct structural characteristics of pygmy modes compared to the giant dipole mode. 

%\clearpage
\begin{acknowledgments}
This work was performed under the auspices of the U.S. Department of Energy by Lawrence Livermore National Laboratory under Contract No. DE-AC52-07NA27344 with partial support from LDRD Projects No. 22-LW-029, 19-ERD-017 and 25-LW-063. 
The work at Brookhaven National Laboratory was sponsored by the Office of Nuclear Physics, Office of Science of the U.S. Department of Energy under Contract No. DE-AC02-98CH10886 with Brookhaven Science Associates, LLC.
We appreciate important contributions to the development of integrated nuclear structure and reaction theory by the late Eric Bauge - he played a central role in establishing the present CEA-LLNL collaboration.
\end{acknowledgments}

\bibliography{myBiB.bib}

\end{document}